# A low cost network of spectrometer radiation detectors based on the ArduSiPM a compact transportable Software/Hardware Data Acquisition system with Arduino DUE

Valerio Bocci, *Member, IEEE*, Giacomo Chiodi, Francesco Iacoangeli, Massimo Nuccetelli, Luigi Recchia

*Abstract*– **The necessity to use Photo Multipliers (PM) as light detector limited in the past the use of crystals in radiation handled device preferring the Geiger approach.**
  **The Silicon Photomultipliers (SiPMs) are very small and cheap, solid photon detectors with good dynamic range and single photon detection capability, they are usable to supersede in some application cumbersome and difficult to use Photo Multipliers (PM). A SiPM can be coupled with a scintillator crystal to build efficient, small and solid radiation detector. A cost effective and easily replicable Hardware software module for SiPM detector readout is made using the ArduSiPM solution [1].**
 **The ArduSiPM is an easily battery operable handled device using an Arduino DUE (an open Software/Hardware board) as processor board and a piggy-back custom designed board (ArduSiPM Shield), the Shield contains all the blocks features to monitor, set and acquire the SiPM using internet network.**

## I. INTRODUCTION

THERE are different materials used as converter from radiation to light. Crystals like CsI(Tl), NaI, Lyso or BGO permit to identify gamma-ray sources using gamma spectroscopy. The detection of thermal-neutron using emitted light can be done by means of a moderator which slows down the neutrons and of a crystals like 6LiI(Eu) Lithium Iodide Europium. The necessity to use Photo Multipliers (PM) as light detector limited in the past the use of crystals in radiation handled device and so as the Geiger approach was preferred. The Silicon Photomultipliers (SiPMs) [2] are very small photon detectors with good dynamic range and single photon detection capability, the small depletion region of SiPMs is not sensible to direct gamma ray respect to other solid-state photons detectors like PINs that can produce false signals.
A SiPM can be coupled with a scintillator crystal so as to build a small and solid radiation detector. The acquisition of the single Silicon Photomultiplier requires different electronics blocks as: preamplifier, discriminator, bias voltage power supply, temperature monitor, Scalers, Analog to Digital Converter and Time to Digital Converter. A cost effective and easily replicable Hardware software module for SiPM detector readout is built using the ArduSiPM solution. The ArduSiPM is an easily battery operable handled device that made out of an Arduino DUE (an open Software/Hardware board)[3] as processor board and a piggy-back custom designed board (ArduSiPM Shield). The Shield contains all the blocks features to monitor, to set and to acquire the SiPM via internet network. The module has a TCP/IP Wi-Fi or Ethernet connection and can send the acquired data like: rate, pulse distribution, timing to a central server or to a single client (like a tablet or a smart phone). A scalable network of multiple Crystal -ArduSiPM based radiation detectors can monitor a local area as a nuclear waste repository or the area around nuclear plants or a large area like a city or an entire country. The cost of a single unit can be very low due to the use of an open platform and System on Chip (SoC) off-the-shelf electronics. The price can be afforded from small institution like schools or home users.

## II. THE ARDUSIPM

The main platform used in the proposed network of radiation detectors is the ArduSiPM [1]. The ARduSiPM project is constituted by three systems : an open software/hardware Arduino DUE Board, a custom designed Arduino Shield, a TCP/IP ethernet or Wifi module , a specific software to control all the function of the shield, a client software.
   Arduino DUE is an open hardware and software development board based on the Atmel SAM3X8E ARM Cortex-M3 CPU. The Arduino DUE is an off the shelf board widespread in the makers community with a free integrated development environment (IDE). The flexibility of the system and the availability of free software sample found applications from sensor control to 3D printer. It is used widespread for didactic platform due to an easy to use program language.
   All the information about the platform are open and the schematics and the development software is available over the internet [2] .
   One of the strengths of the Arduino Board as the first Apple II computer is the possibility to connect custom peripheral card to interface the main processor with sensors and actuators.
   The Arduino DUE can be connected with the external world via a series of connectors present in the edge.

V. Bocci (e-mail:Valerio.Bocci@roma1.infn.it) ,G. Chiodi, F. Iacoangeli, M. Nuccetelli, L. Recchia are with INFN Sezione di Roma , P.le Aldo moro 2, Rome, I-00185, Italy

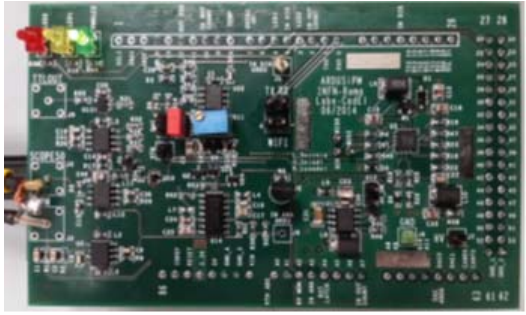

Fig. 1 ArduSiPM Shield

The peripheral card can connected piggyback using the same form factor of the Arduino, in the Arduino world this pluggable boards are called Shields.

There are in the market many shields ranging from motor driver to EMG Shield. Looking to the potentiality of the SAM3X8E ARM Cortex-M3 CPU, we decide to build up a custom shield to acquire Silicon photomultiplier.

The ArduSiPM Shield (Fig. **1** ) is our custom designed board with all electronics interface from Arduino DUE and a SiPM photodetector. The ArduSiPM Shield plugged in the Arduino DUE create an easily transportable system including Front end electronics and data acquisition system.

The global architecture of the system is in Fig. **2**,

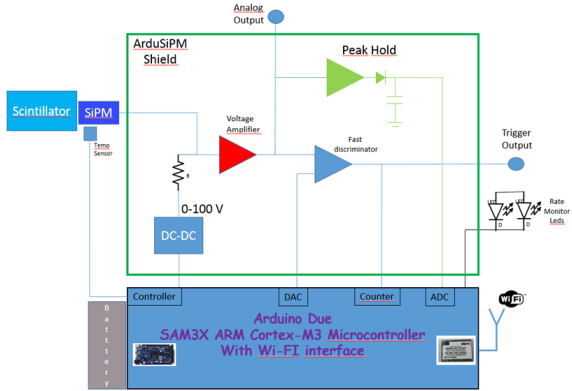

Fig. 2 ArduSiPM Block Diagram

A SiPM and a temperature sensor are connected externally, a digital controlled power supply providing the right voltage to the SiPM, a voltage amplifier, a fast discriminator with a programmable threshold, a peak hold circuit for pulse amplitude, some leds for monitoring, all outputs from analog circuit and digital controls are connected to the Arduino DUE board.
The main components of the ArduSiPM Shield are:

*A. SiPM Power Supply adjustment circuit.*

A DC-DC converter use the Arduino 5 V to generate a supply in the range of 30-100 V. Once we fixed the nominal voltage it is possible to vary around this value of few Volt with an 8 bits resolution. It is possible to use the SiPM voltage to real time compensate the temperature variation typical of this kind of detector.

*B. Analog circuits: Voltage Amplifier and Fast Discriminator*

The voltage amplifier operate a signal conditioning of the SiPM adapting the output to the fast discriminator and the Arduino Analog to Digital converter range.
The Amplifier is fast and linear enough to follow the SiPM response of few ns and to cover all the range of SiPM signal. The noise is less than a single SiPM one pixel signal. For monitoring purpose a replicated output of the amplifier is available as external analog connection.

The SiPM's output Signal is very short (few ns). A fast 7 ns discriminator is used to discriminate the over threshold signals and to count them using Arduino DUE counter. The Threshold value is digitally controlled and his value is monitored using one channel of internal ADC. The width of discriminator is programmable to avoid after pulse counting and to control the death time of the pulse acquisition. A replicated TTL output of the fast discriminator is available as Trigger Output. It can be used as trigger for external acquisition system.

*C. Peak height measurement.*

A precise circuit with fast peak detector is used as peak hold. A sampling comparator features a very short switching charge to measure the short pulses coming out from the SiPM detector when the pulse generate a trigger. The Pulses are stretched over 1 μs to be converted from the 1 MSPS 12 Bits ADC of the SAM3X8E. A programmable digital signal control the fast discharge circuit to reset the circuit of the peak hold after the ADC acquisition, and to rearm the system for a new acquisition

*D. Rate flashing LEDs*

There are two LEDs directly controlled from Arduino DUE, in our software implementation one led flash every over threshold pulse and another every ten pulses in one second windows.

*E. Network interface*

The data elaborated from the ArduSiPM are available through a network interface. This interface can show the data on a tablet or a PC. A PC server can elaborate a network of multiple ArduSiPM so as to monitor a wide area where multiple ArduSiPM are connected.

### III. ARDUSIPM DATA MEASUREMENT.

ArduSiPM is a data acquisition system for SiPM and provide measurements.

*A. ArduSiPM measurements output*

ArduSiPM splits the continuous time in fixed acquisition windows. The default value of each window is one second, in this case the value of number of pulses is coincident with the value in Bequerel of radiation activity.
The pulse parameter: relative time and amplitude are stored in continuous after the arrival of each pulse using interrupt routine. The print of value is done as soon as possible when the network channel is free. The number of arrived pulses is sent at the end of each acquisition window. The buffer act as derandomizer to optimize the data output on the time window. The system is designed to run with negligible dead time.

Any one of this measurement can switch enabled or not enabled:

- The amplitude of each pulse with a 12 bits resolution
- The arrival time respect the start of the time window with a precision up to 25 ns.
- The number of pulse for each time window

The (Fig. **3**) shows the measurements in amplitude A1,A2,A3… and in arrival time t1,t2,t3,…..

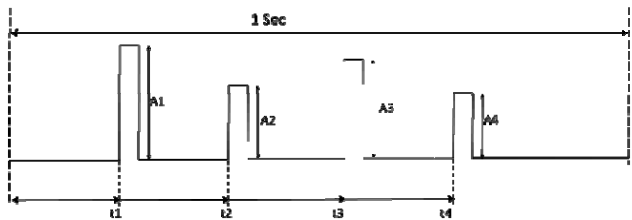

Fig. 3 Measurements in an acquisition window

The output format is readable (Fig. **4**). To optimize the network channel the amplitude and the arrival time are in hexadecimal in variable format with MSB zero suppression and the rate is printed in decimal. The maximum data acquisition rate depends from enabled measurements it is of the order of:
- 20 MHz in rate mode only.
- 4-6 KHz with amplitude value
- 1-2 KHz with arrival time.

```
Cn_y rate:
$10
$50
$244

ADC+Rate:
v1Fv1Dv22v27v_Dv19v20v23v20v1Cv19v1F$12
v18v1Ev1Ev1Bv_9v1Bv29v19v1Av1Dv1Bv1Dv2Av18v1B$15
v15v20v21v21v_Dv1Fv1Av1Av1A$9
v19v17v1Bv18v`Cv1Dv1D$7

TDC+ADC+RATE:
taedvataf0v7tv9v3$3
```
Legend:
vXXX ADC Value in HEX MSB zero suppressed
XXXXXXXX TDC value in HEX MSB zero suppressed
$XXX rate in Hz

Fig. 4 Example of data stream comings out from the ArduSiPM

### B. Analog chain linearity.

We test our system to understand the performance. The digital part gives a perfect correspondence in rate and time measurements by construction, the analog chain: voltage Amplifier, peak hold circuit and ADC show a very good linearity as we measure in Fig. **5**.

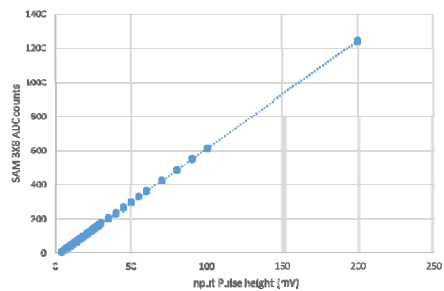

Fig. 5 test of analog chain linearity injecting a calibration pulse and reading the ADC value

### IV. COMMERCIAL USABLE CRYSTALS

The SiPM of the acquisition board can be coupled with a scintillator crystal to build efficient, small and solid radiation detector. The first application ArduSiPM was used to measure beta ray medical tracer using para-Therphenil scintillator [4] and a BC408 plastic scintillator so as to monitor particle beam. The gamma radiation monitoring is an interesting field of application for the ArduSiPM. The gamma radiation can travel distant from the source and can be used to identify the type of radiatioactive sources using the gamma spectra analysis [6]. The first component to convert the gamma ray to light detectable from the SiPM is a crystal there are different kind of crystals for the scope. Scintillation materials can be small, low-cost, and efficient for gamma detection, can operate at room temperature, and are capable of being used in spectroscopy systems. The volume of crystal necessary to detect gamma ray can be in the order of one cube centimeter. The **Error! Reference source not found.** most common crystal usable with ArduSiPM.

The common NaI(Tl) can be used but in a system that can work in open field the Higroscopicity of the material is a contraindication. The CsI(Tl) [5] is good especially for the very good Light Yeld even if it has slight Higroscopicity. The BGO and the LSO (LYSO)[7] cristals can be used even if they have a smaller Light Yeld. All the listed crystals are suitable for the detection of Gamma ray from hundreds Kev to some MeV using a crystal size of the order of one cube centimeter.

### V. USE OF ARDUSIPM AS NEUTRON DETECTOR

One of the use of the ArduSiPM can be the detection of Neutrons. The common neutron's detectors are complex and heavy systems, mainly based on light conversion and PM readout.
A small detector can be built to convert neutron detection in light. One proved technics is to use a LiI crystal viewed by a SiPM. An interesting ready to use design for the detector is published [8]. It consists on a detector of 23 mm diameter LiI crystal viewed by a 14x14mm square SiPM, a 10mm thick

HDPE moderator in front of the 3mm crystal and an acrylic light-guide behind it.

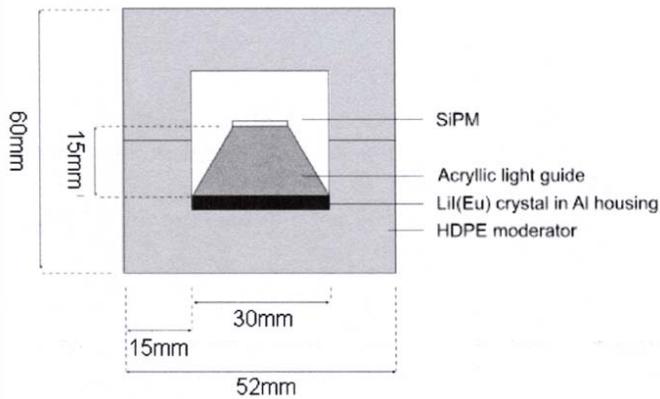

Fig. 6 A small neutron detector using a SiPM as proposed in [8].

The neutron moderation is provided by high-density polyethylene (HDPE), then the slow neutrons interact with the LiI crystal. Lithium iodide, when suitably activated, scintillate under slow neutrons' irradiation as a result of the 6Li(n,a)3H reaction in which the a-particle and triton share an energy of 4.79 MeV. The scintillation efficiency is 11000 ph/MeV. The Acrylic light guide conveys the light to the SiPM detector. This kind of detector can be easily read from the ArduSiPM device and can meet the DNDO (Domestic Nuclear Detection Office) specification for a handheld radiation detector.

## VI. USE OF ARDUSIPM AS A CHEAP STANDALONE GAMMA SPECTROMETER WITH TABLET/ PC INTERFACE

The communication protocol of the ArduSiPM first version use a light textual based interface and can be used like a terminal.
This is a powerful for expert users but can be heavy for inexpert. The lack of Graphical user interface can be brightly resolved using as client a PC or a Tablet. We developed an Android App that can control and display data in raw or in graphical way after post processing (Fig. 7).

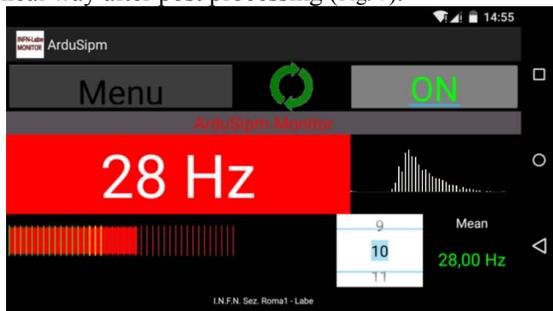

Fig. 7 Screenshot of the ArduSiPM control APP.

## VII. USE OF MULTIPLE ARDUSIPM IN A RADIOLOGY ENVIRONMENT MONITOR NETWORK

The very low cost of the spectrometer built around ArduSiPM permits to think of a wide use of multiple system in a nuclear waste disposal or to use it as element in a wide area network of nuclear radiation monitoring involving schools or single people in Crowd-Sourcing. There are different activities using radioactive materials: nuclear power generation, defense, medicine, and scientific research. Any activity that produces or uses radioactive materials generates radioactive waste by products. Radioactive waste can be in gas, liquid or solid form, and its level of radioactivity can vary. Designs for new disposal facilities and disposal methods must meet environmental protection and pollution prevention standards that are stricter than were foreseen at the beginning of the atomic age. The availability of a multiple spectrometers around the nuclear waste disposal can give a real time monitoring of the level of radiation with minimal human intervention.

After the nuclear accident in Fukushima, Japan, in 2011 the public demand of radiation monitor increase considerably. Many people were unsure what level of radiation they were being exposed. There are projects like EURDEP [9] (EUropean Radiological Data Exchange Platform Fig. 8) that show radioactivity and emergency preparedness in the Europe area. EURDEP makes radiation dose rate data from 39 organizations in 37 European countries, there are 4500 automatic stations available on an hourly basis and in addition from some 100 air concentration monitoring stations on a daily basis during an emergency, as well as under normal conditions. The freely accessible Public map allows the public to view the European monitoring data.

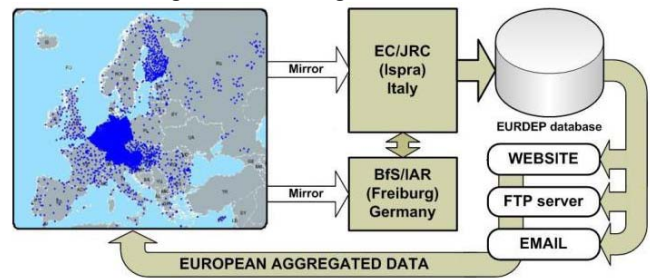

Fig. 8 EURDEP (European Radiological **Data** Exchange Platform)

The EURDEP network use as standard exchange format IRIX (International Radiation Information eXchange). The IRIX is an xml-based format standard for data exchange that has been developed under the IAEA action plan, in closed cooperation with the EC. The IRIX is well designed to exchange of environmental radiation data include appropriate metadata with the results of monitoring. The ArduSiPM can easily output the data in IRIX format and can be integrated in EURDEP or other radiation network like RadiationNetwork.com.

## VIII. CONCLUSIONS

The ArduSiPM coupled with an appropriate scintillator material can be used as a cheap gamma ray spectrometer or neutron detector. Multiple ArduSiPM device can be used as building block of a network of sensors for a nuclear waste disposal or a wide area radiation monitor system.

The price can be afforded from small institution like schools or home users to build up a Smart Urban Crowd-Sensing.

ACKNOWLEDGMENT

We want to thanks Francesco Di Lorenzo for one of the first Schematic layout of the ArduSiPM, Riccardo Lunadei for the PCB layout.